# Precise Determination of the Crystallographic Orientations in Single ZnS Nanowires by Second-Harmonic Generation Microscopy

Hongbo Hu,[1] Kai Wang,[1,*] Hua Long,[1] Weiwei Liu,[1] Bing Wang,[1,*] Peixiang Lu,[1,2,*]

[1]Wuhan National Laboratory for Optoelectronics and School of Physics, Huazhong University of Science and Technology, Wuhan 430074, China

[2]Laboratory for Optical Information Technology, Wuhan Institute of Technology, Wuhan 430205, China

*Corresponding authors: kale_wong@hust.edu.cn (KW), wangbing@hust.edu.cn (BW), lupeixiang@hust.edu.cn (PXL)

ABSTRACT

We report on the systematical study of the second-harmonic generation (SHG) in single zinc sulfide nanowires (ZnS NWs). The high quality ZnS NWs with round cross-section were fabricated by chemical vapor deposition method. The transmission electron microscopy images show that the actual growth-axis has a deviation angle of $0^o$~$20^o$ with the preferential growth direction [120], which leads to the various polarization-dependent SHG response patterns in different individual ZnS NWs. The SHG response is quite sensitive to the orientations of *c*-axis as well as the (100) and (010) crystal-axis of ZnS NWs, thus all the three crystal-axis orientations of ZnS NWs are precisely determined by the SHG method. A high SHG conversion efficiency of $7\times10^{-6}$ is obtained in single

ZnS NWs, which shows potential applications in nanoscale ultraviolet light source, nonlinear optical microscopy and nanophotonic devices.

Recently, the nonlinear optical properties of semiconductor nanowires (NWs) have attracted widespread interest for their potential applications in nanophotonic devices,[1] all-optical switching,[2] optical probe[3] and life science.[4-6] Second-harmonic generation (SHG) is a second-order nonlinear optical process that generates photon with twice the fundamental frequency of the pumping laser. It provides a convenient approach to obtaining ultraviolet light using a near-infrared laser. As previously reported, SHG in NWs of several materials such as alkaline niobates,[7] ZnO,[8] ZnTe,[9] CdS,[10] GaAs,[11] GaN[12] and GaP[13] have been demonstrated to be highly efficient and controllable. In particular, SHG is quite sensitive to the polarization angle of the pumping light with respect to the crystallographic orientation of materials. The measured polarization-dependent SHG responses can be used to determine the crystallographic orientations of nanostructures.[12, 14, 15] It offers an all-optical method for in situ determination of crystal orientations without sample damages or special environmental requirements, which can be an alternative to electron diffraction methods. However, previous studies mostly focused on the influence of the *c*-axis orientation of nanostructures on the SHG response.[12, 15, 16] The effect of the *a*- and *b*-axis orientations is usually ignored, which may omit some important detailed information on the SHG in semiconductor NWs.

Zinc sulfide (ZnS) is an important semiconductor with merits of chemical stability and low toxicity, which has a wide range of industrial applications.[17, 18] Specifically, ZnS is a direct wide band-gap semiconductor (3.72 and 3.77eV for cubic and hexagonal structure, respectively, at 300 K) with a large exciton binding energy of 40 meV.[19] Besides, ZnS is transparent in the near-ultraviolet, visible and near-infrared regions with large refractive indices ($n$=2.3~2.5),[20] which can be an ideal candidate for the high efficient nonlinear-optical devices,[21] nanoscale near-ultraviolet light source[22] and sensors.[23] Thus, it is valuable to systematically study the SHG in single ZnS NWs with the precise determination of the NW crystallographic orientations. It not only provides a better understanding of the polarization-dependent SHG properties in NWs, but also offers important experimental information on the NW-based polarization devices and the nanoscale ultraviolet light source.

In this letter, we report the first demonstration on the polarization-dependent SHG in single ZnS NWs with a high SHG conversion efficiency of $7\times10^{-6}$. The high-resolution transmission electron microscopy (HRTEM) images show that the actual growth-axis of ZnS NWs has a deviation angle of $0^\circ$~$20^\circ$ with the preferential growth direction [120], resulting in various SHG response patterns in different individual ZnS NWs. It is found that not only the the $c$-axis, but also the other two crystal axes of ZnS NWs have influence on the SHG radiation patterns.

The ZnS NWs were fabricated in a horizontal tube furnace by a conventional CVD method. ZnS powder (99.99%) was loaded in a quartz boat and placed in the center of a quartz tube. Several slices of single-crystal Si (001) wafer covered with Au catalyst nanoparticles (30 nm) were placed downstream of the source materials, acting as the deposition substrates. Prior to heating, the quartz tube was purged with high-purity argon (Ar) gas for one hour. After that, the source was heated to 1050 ºC at a rate of 30 ºC/min, and maintained at the top temperature for one hour. During this process, the quartz tube was kept at a pressure of 0.03 MPa with the high-purity Ar gas (carrier gas) introduced at a constant flow rate of 80 sccm. After the system was cooled naturally, white wool-like products were deposited on the silicon substrates. Single ZnS NWs were then transferred onto a quartz substrate for the further optical measurement.

The products were characterized by X-ray diffraction (XRD, X'Pert PRO, PANalytical B.V., Netherlands), field-emission scanning electron microscope (FESEM, Nova NanoSEM 450) and high resolution transmission electron microscopy (HRTEM, Tecnai G220). A typical SEM image of the as-synthesized ZnS NWs on silicon substrates is shown in **Figure 1a**. The diameters of the NWs range from 300 nm to 800 nm and the lengths are up to tens of micrometers. **Figure 1b** shows a single ZnS NW lying on a quartz substrate with a round cross-section and uniform diameter. The XRD pattern of the sample shown in **Figure 1c** is in good agreement with JCPDS card no. 36-1450, indicating a hexagonal wurtzite structure of the ZnS NWs with the lattice constants of $a=b=0.38$ nm and $c=0.63$ nm. The sharp peaks without shifts suggest the high phase purity of the sample. **Figures 1d-1f** present the HRTEM images and corresponding SAED patterns of three different ZnS NWs, which further confirm the wurtzite crystal structure and the high crystalline quality of the sample. Importantly, although the ZnS NWs preferentially grow along the crystal direction [120], the actual growth-axis is not accurately fixed to it. The deviation angles are measured to be 20º, 11º and 0º for the samples in **Figures 1d-1f**, respectively. Since the crystal direction [120] is perpendicular to the crystal direction [001] ($c$-axis), the $c$-axis of the ZnS NWs is not perpendicular to the actual growth-axis, but exhibits deviation angles of 70º~90º. This deviation in growth directions may be caused by the vapor-liquid-solid (VLS) growth mechanism in the ZnS NWs.[18, 24] Since the single ZnS NWs with round cross-section are randomly distributed on a quartz substrate, the observed deviation of the growth-axis can lead to a significant variation in the polarization-dependent SHG responses from different individual ZnS NWs.

The SHG experiment was carried out by a conventional confocal microscope system at room temperature. **Figure 2a** shows the schematic sketch of the experimental setup. A mode-locked Ti-sapphire femtosecond laser system (Tsunami, Spectra-Physics, ~810 nm, 50 fs and 80 MHz) was used as the pumping source. The intensity of the pumping laser beam was adjusted by combining a half-wave plate ($A_1$) and a polarizing beam splitter (**B**). In addition, the polarization direction of the pumping laser was controlled by another half-wave plate ($A_2$). A microscope objective (Olympus, 0.65 NA, 40X) focused the pumping laser onto the sample with a focal spot diameter of ~4 μm. The transmitted SHG signal originated from ZnS NWs was collected by another identical objective, and subsequently imported to a CCD or spectrometer (Princeton Instruments Acton 2500i with Pixis

CCD camera). A 750-nm short-pass filter (**C**) filtered out the pumping laser. The polarization-dependent SHG response of the single ZnS NWs was measured by rotating the polarization direction of the pumping laser with **A₂**. As shown in **Figure 2b**, the pumping laser propagates along the Z-axis, while a single ZnS NW lies along the X-axis. The electric-field of the linearly polarized pumping laser makes a controllably variable angle $\theta$ with respect to the growth-axis of the ZnS NW (namely X-axis). As illustrated in **Figure 2c**, the crystal frame ($x_c y_c z_c$ axes) for ZnS crystals is assumed arbitrary orientation that is defined by three Euler angles ($\varphi$, $\gamma$, $\omega$) in the laboratory frame (XYZ axes). Specifically, $\varphi$ is the angle between $z_c$-axis and Z-axis, $\gamma$ is the angle between X-axis and the $z_c$-axis's projection on XY-plane, and $\omega$ is the angle between $x_c$-axis and the intersection line of XY- and $x_c y_c$-planes. It should be noted that the orientation of the $z_c$-axis is consistent with the [001] *c*-axis of the ZnS NWs, which is defined by the angles $\varphi$ and $\gamma$. Besides, the angle between $x_c$-axis and $y_c$-axis for the wurtzite crystal frame is 120°. In order to avoid ambiguity, these Euler angles are restricted within the range of (0°, 90°) for $\varphi$, (-90°, 90°) for $\gamma$, and (-60°, 120°) for $\omega$, respectively.

The sample for the SHG measurement was the isolated single ZnS NW on a quartz substrate to avoid the mutual effect between different ZnS NWs. The signal spectrum in **Figure 2d** presents a strong peak at 405 nm, which is exactly the frequency doubling signal of the pumping laser. Furthermore, no peaks from defect-related photoluminescence emissions[25] can be observed, indicating the high crystalline quality of the ZnS NWs. **Figure 2e** shows the signal intensity as a function of the pumping laser power. The quadratic dependency indicates the signal is generated from a second-order nonlinear process. By combining the spectrum in **Figure 2d**, it is concluded that the signal is ascribed to the SHG process in a single ZnS NW. The polarization of the SHG signals was studied by rotating a Glan Prism in front of the spectrometer. The results indicate that the polarization of the measured SHG signal does not keep linearly polarized but varies with the orientation of ZnS crystal and the polarization direction of pumping laser. (see the Supporting Information.) The hexagonal wurtzite ZnS belongs to non-centrosymmetric crystal class 6mm, which allows for a highly efficient SHG.[26, 27] An estimation of the average SHG conversion efficiency is up to 7×10$^{-6}$ (see the Supporting Information), which shows a potential for practical applications.

The polarization-dependent SHG response is sensitive to the crystallographic orientation. For theoretical considerations (see the Supporting Information), it is worth noting that the diameter of ZnS NWs is much smaller than the focus spot of the pumping laser (~4 μm). By further ignoring the materials birefringence, the electric field **$E_i$** at fundamental frequency inside the nanowire can be regarded as a uniform field. To calculate the second-order polarization induced by the pumping laser, the electric field **$E_i$** has to be firstly projected onto the crystal coordinate system, namely **$E_{cx}$**, **$E_{cy}$** and **$E_{cz}$** on the $x_c, y_c$ and $z_c$ axes in crystal frame shown in **Figure 2c**, respectively. According to Euler's rotation theorem, electric field components of the fundamental frequency in the crystal frame, **$E_{cx}$**, **$E_{cy}$** and **$E_{cz}$** are defined by:

$$\begin{bmatrix} E_{cx} - E_{cy}/2 \\ \sqrt{3} \cdot E_{cy}/2 \\ E_{cz} \end{bmatrix} = \begin{bmatrix} \cos(\gamma - \theta)\cos(\varphi)\sin(\omega) + \sin(\gamma - \theta)\cos(\omega) \\ \cos(\gamma - \theta)\cos(\varphi)\cos(\omega) - \sin(\gamma - \theta)\sin(\omega) \\ \cos(\gamma - \theta)\sin(\varphi) \end{bmatrix} \cdot E_i \quad (1)$$

The wurtzite ZnS belongs to *6mm* crystal class and has nonzero second-order nonlinear susceptibility components, $d_{15}$=8.0 pm/V, $d_{31}$=8.1 pm/V and $d_{33}$= -17 pm/V in the electrical-dipole approximation.[28] Since ZnS lacks inversion symmetry, higher-order multipole processes, such as the electric quadrupole processes, are much weaker than the dipole allowed process.[29] Thus, our analysis only considers the dipole processes. The SHG polarization components (**$P_{cx}$**, **$P_{cy}$** and **$P_{cz}$**) along the three crystal axes are related to **$E_{cx}$**, **$E_{cy}$** and **$E_{cz}$** by:

$$\begin{bmatrix} P_{cx} \\ P_{cy} \\ P_{cz} \end{bmatrix} = 2\varepsilon_0 \begin{bmatrix} 0 & 0 & 0 & 0 & d_{15} & 0 \\ 0 & 0 & 0 & d_{15} & 0 & 0 \\ d_{31} & d_{31} & d_{33} & 0 & 0 & 0 \end{bmatrix} \cdot \begin{bmatrix} E_{cx}^2 \\ E_{cy}^2 \\ E_{cz}^2 \\ 2E_{cy}E_{cz} \\ 2E_{cx}E_{cz} \\ 2E_{cx}E_{cy} \end{bmatrix} \quad (2)$$

The second-order polarization components in the ZnS NWs can be regarded as electrical dipoles that oscillate at the SHG frequency. Since electrical dipole does not radiate in a simple spherical wave, it is necessary to consider the collection efficiency of the objective for an accurate estimation of the SHG responses. The collected SHG power $P_{SHG}$ that is radiated from the three dipole antennas is defined as[30]

$$P_{SHG} = \frac{ck^4V^2}{12\pi\varepsilon_0}\left(\eta_{cx}|P_{cx}|^2 + \eta_{cy}|P_{cy}|^2 + \eta_{cz}|P_{cz}|^2\right) \qquad (3)$$

where $c$ is the speed of light, $k$ is the wave number at the SHG frequency, $V$ is the volume of ZnS with second-order polarization induced by the fundamental optical field, $\varepsilon_0$ is the vacuum permittivity, $\eta_{cx}$, $\eta_{cy}$ and $\eta_{cz}$ are the objective collection efficiency for the three SHG polarizations $P_{cx}$, $P_{cy}$ and $P_{cz}$, respectively. Since the second-order polarization components ($P_{cx}$, $P_{cy}$ and $P_{cz}$) are all as the function of polarized angle $\theta$, the relation between SHG signal intensity $I$ and $\theta$, can be obtained from **Equation 3**.

The crystallographic orientations of the ZnS NWs can be determined by fitting the measured SHG response as a function of $\theta$ with **Equation 3**. **Figures 3a** and **3d** show the polarization-dependent SHG response of two different single ZnS NWs. The black dots present the experimental data, and the solid curves indicate the theoretical fits. The polar plots present a symmetrical shape of two-lobe. As can be seen, the fitted curves agree well with the experimental data. Note that the coefficient of determination ($R^2$) of the fitted curves are all larger than 0.97, indicating a reasonable estimation for the polarization-dependent SHG by the theoretical models. Euler angles for the two samples in **Figures 3a** and **3d** are determined to be ($\varphi = 85°$, $\gamma = -69°$ and $\omega = 4°$) and ($\varphi = 58°$, $\gamma = -67°$ and $\omega = -10°$), respectively. In order to investigate the influence of angle $\gamma$ on the SHG responses patterns, the theoretical calculation results in **Figures 3b** and **3c** are plotted by just changing $\gamma$ while keeping $\varphi$ and $\omega$ the same as that in **Figure 3a**. As shown in **Figures 3a-3c**, it can be seen that the change of $\gamma$ just rotates the two-lobe pattern and has no influence on the pattern shape. Thus, the pattern shape is determined by $\varphi$ and $\omega$ while $\gamma$ contributes to the orentation of the two-lobe patterns. This conclusion is further supported by the results shown in **Figures 3d-3f**. Moreover, the results shown in **Figures 3a-3c** are all plotted at $\varphi = 85°$. It's able to speculate that when $\varphi$ is 90°, $\gamma$ equals the angle between the long axis of the two-lobe pattern and the X-axis. As a matter of fact, the strongest SHG intensity occurs as the electric field of the pumping laser is parallel to the *c*-axis of the NWs, which is ascribed to the largest second-order nonlinear coefficient component ($d_{33}$) of ZnS. The influence of $\gamma$ on the SHG responses can be explained by the projections of three polarization components along crystal frame on XY-plane. Obviously, $\gamma$ does not change the magnitude of $\boldsymbol{P}_{cx}$, $\boldsymbol{P}_{cy}$ and $\boldsymbol{P}_{cz}$'s projection on XY-plane under a certain ($\varphi$, $\omega$) pair. Instead, it just rotates these three projections on XY-plane, which results in a rotation of the two-lobe pattern with shape unchanged.

The influence of angle $\omega$ on the SHG response is also studied. **Figures 4a** and **4c** show the polarization-dependent SHG polar plots of two single ZnS NWs, which exhibit the similar two-lobe patterns. The distinction between them is the different locations of the small protuberances. The peaks of small protuberances in **Figures 4a** and **4c** are located at $\theta=165º$ and $\theta=150º$, respectively. According to the theoretical fittings, the Euler angles for the samples in **Figures 4a** and **4c** are determined to be ($\varphi=76º$, $\gamma=70º$ and $\omega=5º$) and ($\varphi=76º$, $\gamma=68º$ and $\omega=54º$), respectively. Both samples have the similar angles of $\varphi$ and $\gamma$, but a rather different $\omega$. It is worth noting that the values of $\omega$ in both patterns are almost symmetric with respect to 30º. **Figure 4b** shows the calculated SHG pattern at $\omega=30º$ while keeping $\varphi$ and $\gamma$ the same as that in **Figure 4a**. Generally, it presents a similar two-lobe pattern to that in **Figures 4a** and **4c**, but is with the small protuberances arising exactly along the direction ($\theta=157.5º$) perpendicular to the long-axis of the pattern. Therefore, the significant variation of the small protuberances in these patterns can be attributed to the influence of angle $\omega$. As $\omega$ varies with respect to 30º, the peaks of the small protuberances deviate accordingly. Since $\omega$ are nearly symmetric about 30º in **Figures 4a** and **4c**, the deviation of the small protuberances present a reciprocal process according to the pattern in **Figure 4b**. As illustrated in **Equation 2**, $\boldsymbol{P}_{cz}$ is the biggest component of the second-order polarization, which contributes to the main part of the lobe patterns, and the components $\boldsymbol{P}_{cx}$ and $\boldsymbol{P}_{cy}$ determine the details such as the small protuberances in the patterns. Note that when $\omega$ equals 30º, the $\boldsymbol{x}_c$- and $\boldsymbol{y}_c$-axis are symmetrical about the plane formed by $\boldsymbol{z}_c$- and *c'*-axis shown in **Figure 2c.** The variation of $\omega$ with keeping $\varphi$ and $\gamma$ as constants can be regarded as the rotation of the $x_c$- and $y_c$-axis around a fixed $z_c$-axis. Therefore, it just changes the projections of $x_c$- and $y_c$-axis on XY-plane and leads to the non-negligible details in SHG response patterns.

**Figure 5** shows the influence of angle $\varphi$ on the polarization-dependent SHG responses. The ZnS NW in **Figure 5a** was found oriented at ($\varphi$=76º, $\gamma$=80º and $\omega$=30º). **Figures 5b** and **5c** plot the theoretical calculation patterns by just changing the value of $\varphi$ while keeping $\gamma$ and $\omega$ the same as the parameters in **Figure 5a**. As the decrease of $\varphi$, the SHG pattern shrinks and the small protuberances in **Figure 5a** gradually disappear. In fact, the variation of angle $\varphi$ would change the projections of all three axes on XY-plane, especially for the $z_c$-axis. It changes the magnitude of all three second-order polarization components. Therefore, angle $\varphi$ has an influence on the total shape of two-lobe patterns. The results show that the nearer $\varphi$ approximates to 0º, the smaller the two-lobe pattern is. Hence, the ZnS NWs with the $z_c$-axis parallel to the optical axis (Z-axis) possesses the lowest SHG conversion efficiency. Additionally, the influence of angle $\varphi$ on the pattern will be effected by the $\omega$ values. As shown in **Figures 5d**-**5f** that $\omega$ equals 71º, the SHG patterns also shrink as $\varphi$ decreases, but the evolution in details is different from that in **Figures 5a**-**5c**. In fact, under different values of $\omega$, the three second-order polarizations show a different relatively change rates as $\varphi$ varies. Therefore, the related evolution processes of the SHG patterns are different as $\varphi$ decreases.

To evaluate the accuracy of this method for diagnosing crystallographic orientation, the SHG fitting results are further verified by the HRTEM results of the ZnS NWs. For purpose of relating the planar TEM patterns to the tridimensional crystal orientations measured by SHG, we discuss the issue in two aspects: angle $\beta$ (the angle between $z_c$-axis and the growth axis of nanowire) and angle $\omega$. Generally, $\beta$ varies in a range of 70º~90º in most of ZnS NWs according to the HRTEM images. The angle $\beta$ measured in SHG, is determined by angle $\varphi$ and $\gamma$. In geometrical theories, the relation between $\beta$ and $\varphi$, $\gamma$ can be written as $\cos(\beta)=\sin(\varphi)\cos(\gamma)$. For example, the ZnS NW in **Figure 4a** with $\varphi$=76º and $\gamma$=70º, angle $\beta$ is calculated to be 71º that is in good agreement with the HRTEM result of 70º in **Figure 1d**. It is pointed out that the agreement holds for all the ZnS NWs in our experiment. Then angle $\omega$ is considered. For simplicity, we assume the growth-axis is in the plane formed by the crystal directions [120] and [001]. Staying with the ZnS NWs in **Figure 1d**, as the NW rotates around its growth-axis to make $\varphi$ and $\gamma$ equal the value of **Figure 4a** ($\varphi=76°$, $\gamma=70°$), the value of $\omega$ calculated by geometrical relationship is 4° (see the Supporting Information), which are quite approximate to the value ($\omega=5°$) fitted by SHG in **Figure 4a**. By the examinations, it is found that most of the fitted values of $\omega$ by SHG response satisfy the aforementioned condition (the growth axis is in the plane formed by crystal directions [120] and [001]). However, the sample in **Figure 5a** ($\varphi=76°$, $\gamma=80°$ and $\omega=30°$) is an exception that the angle $\omega$ is calculated to be 3° by the above geometrical relationship. Actually, it is not an experiment error but ascribed to a more complex geometrical condition. Specifically, for some samples that the NW growth-axis is not in the plane formed by the crystal directions [120] and [001], as shown in **Figure 1e** where the $z_c$-axis [001] is not in the paper plane. The geometrical method for evaluating $\omega$ should be modified under the condition, and the fitting results of **Figure 5a** are proved to be reasonable (see the Supporting Information). Although the crystallgraphic determinations based on the SHG analysis are proved to be effective and accurate, there is a special situation when $\varphi$ equals 90°. Specifically, the SHG patterns for the Euler-angle of (90°, $\gamma_0$ and $\omega_0$) and (90°, $\gamma_0$ and 60°-$\omega_0$) would be the same. As a result, $\omega$ is ambiguous in this situation. Anyway, this theoretical model for the far-field polarization-dependent SHG repsonses provides an effective and convenient method for precise determinations of crystallographic orientations of materials.

As discussed, the SHG properties in ZnS NWs show to be efficient and polarization sensitive. In previous reports, the growth process of the ZnS nanostructures is highly controllable that several kinds of nanostructures have been fabricated, such as quantum dots, nanorods, nanoribbons et al.[31-34] Thus, ZnS nanostructures present advantages in the coherent and tunable nanoscale light sources and nanoprobes especially for the ultraviolet region. In particular, the ZnS-based heterostructures, such as core/shell particles,[35] co-axis NWs[36, 37] can be conveniently obtained by conventional chemical methods. The SHG conversion efficiency can be further enhanced with a proper optical confinement in the ZnS-based heterostructures, which may be planned for a further work.[38]

In summary, we report on the polarization-dependent SHG response of single ZnS NWs. The polarization-dependent SHG response patterns were mainly attributed to the *c*-axis orientations of the ZnS NWs. Meanwhile, for the first time, the orientations of the other two crystal axes were also found to have influence on the SHG patterns. Based on the SHG method, all the three crystal axes orientations of ZnS NWs were precisely determined. It provides a convenient all-optical method for in situ determination of crystal orientations without sample damages. In addition, a high SHG conversion efficiency of $7\times10^{-6}$ was achieved in the single ZnS NWs, indicating potential applications in the nanoscale ultraviolet source, nonlinear optical microscopy and nanophotonic devices.

ACKNOWLEDGMENT

This work was supported by the 973 Program under grant 2014CB921301 and National Natural Science Foundation of China (11204097), the Doctoral fund of Ministry of Education of China under Grant No. 20130142110078. Special thanks to the Analytical and Testing Center of HUST and the Center of Micro-Fabrication and Characterization (CMFC) of WNLO for using their facilities.

SUPPORTING INFORMATION PARAGRAPH: Estimation of the second-harmonic generation (SHG) conversion efficiency; theoretical analysis of the polarization-dependent SHG response in ZnS NWs; the coherence of the SHG signal; variation of ω as ZnS NW rotates around its growth axis. This material is available free of charge via the Internet at http://pubs.acs.org.

FIGURE CAPTIONS

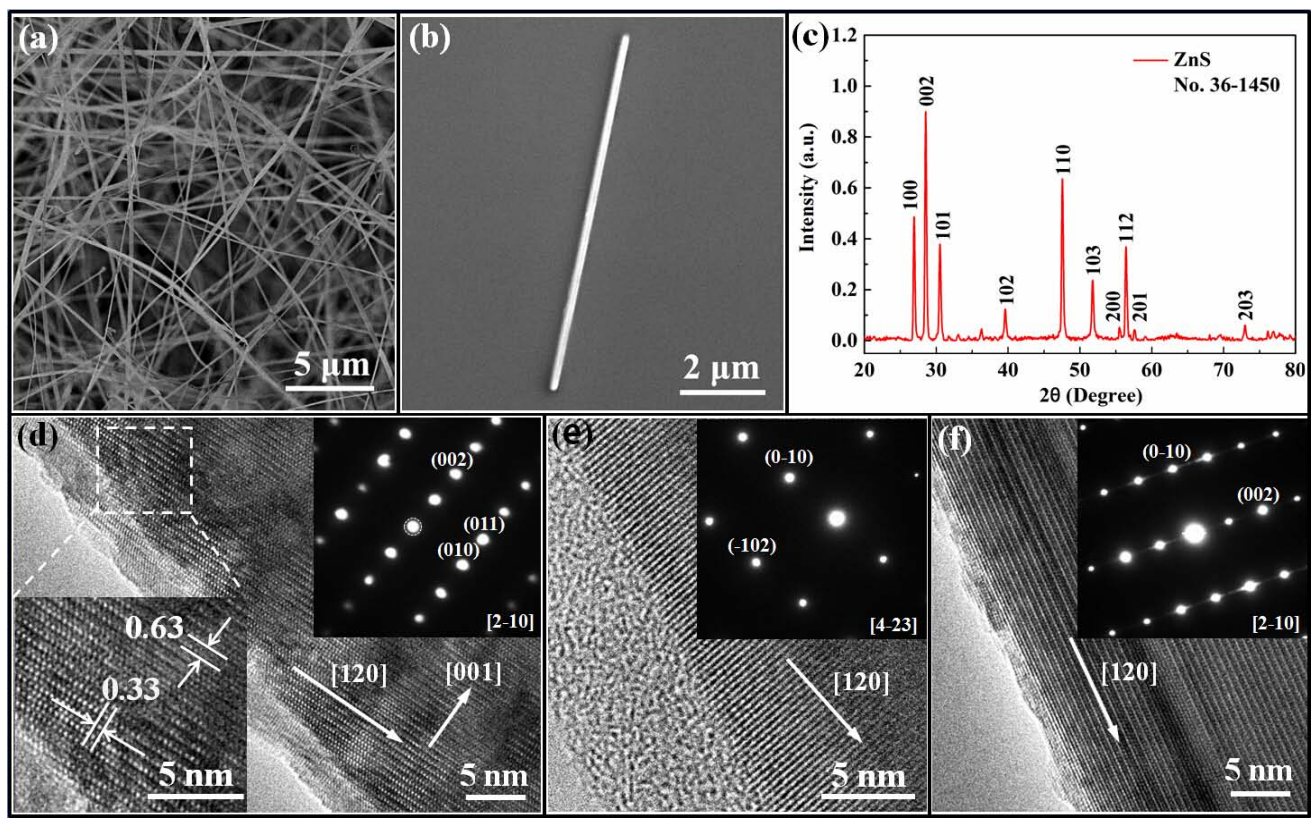

**Figure 1.** **(a)** The SEM image of the ZnS NWs grown on a silicon substrate. **(b)** The SEM image of a single ZnS NW on a quartz substrate. **(c)** The XRD pattern of the ZnS NWs. **(d)-(f)** The HRTEM images and corresponding SAED patterns of three different single ZnS NWs with diverse growth directions. The deviation angles between the growth-axis and crystal direction [120] are **(d)** 20º, **(e)** 11º and **(f)** 0º, respectively.

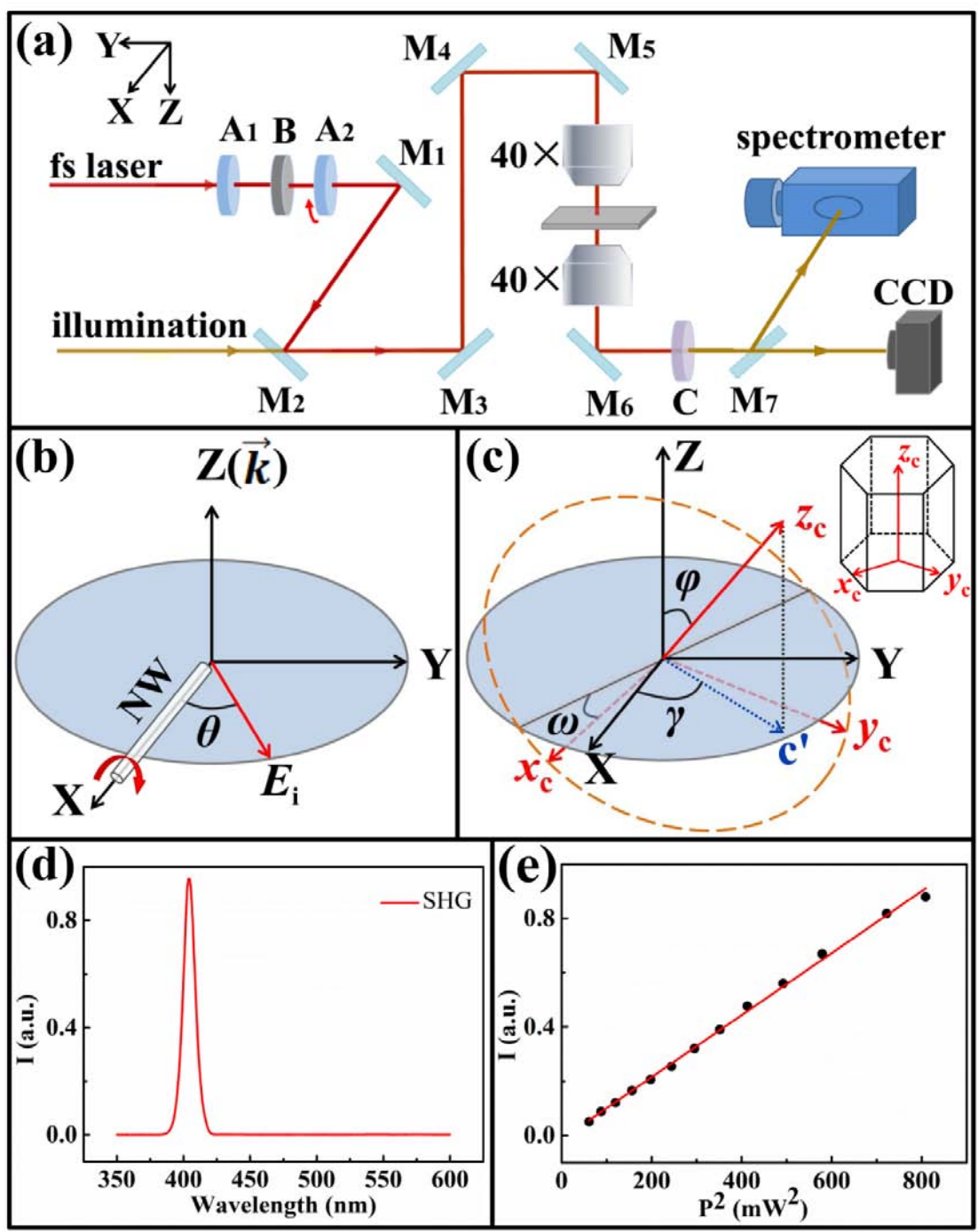

**Figure 2.** **(a)** The schematic sketch of the experiment setup for SHG measurements on single ZnS NWs. **A1**, **A2**: half-wave plates at 800 nm; **B**: polarizing beam splitter; **C**: 750-nm short pass filter. **M2**: an ultrafast mirror with the working wavelength of 700 nm to 930 nm, and it is transparent to most of visible light. The pumping laser was focused by a 40x objective (Olympus, NA=0.65), and the transmitted SHG signal was collected by another 40x objective. **(b-c)** Geometries of the lab frame (XYZ) and the crystal frame ($x_c y_c z_c$). **(b)** The linearly polarized pumping laser propagates along Z-axis. The optical electric field $E_i$ of the pumping laser is in the XY-plane with a variable

DOI: 10.1021/acs.nanolett.5b00607

angle $\theta$ to the NW growth-axis. **(c)** The relative position of the crystal frame in the lab frame. Specifically, $\varphi$ is the angle between $z_c$-axis and Z-axis, $\gamma$ is the angle between X-axis and the $z_c$-axis's projection (c') on XY-plane, and $\omega$ is the angle between $x_c$-axis and the intersection line of XY- and $x_c y_c$-planes. The orientation of the crystal $z_c$-axis is consistent with the **c**-axis of the ZnS NWs, which is defined by the angles $\varphi$ and $\gamma$. **(d)** Spectrum of the SHG signal originated from a single ZnS NW under a pumping power of 6 mW. **(e)** Measured SHG signal intensity $I$ as a function of the square of the pumping power $P^2$.

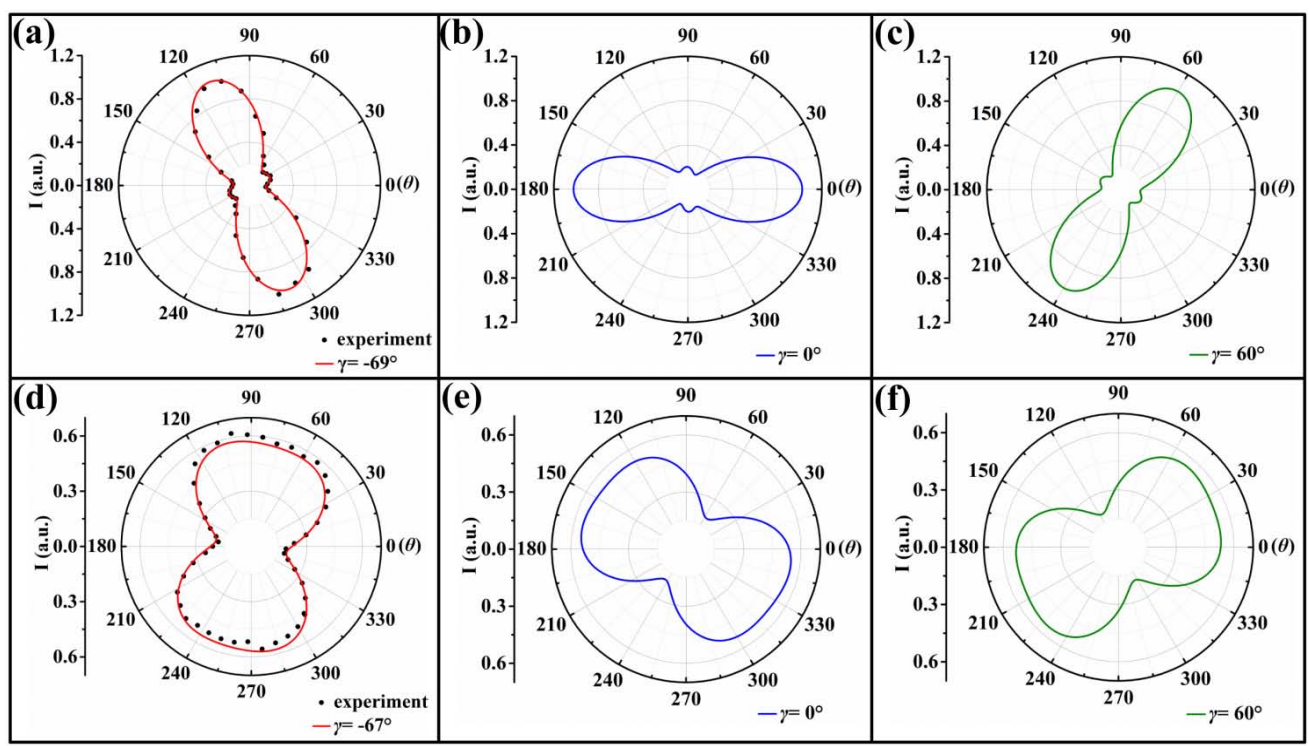

**Figure 3.** Polar plots of the measured SHG intensity as a function of the polarization angle $\theta$. The black dots present the experimental data, and the solid curves indicate the theoretical fits. **(a)** The crystal orientation of ZnS NW is determined to be $\varphi=85°$, $\gamma= -69°$ and $\omega=4°$. **(b)** and **(c)** the related theoretical calculations by just changing the value of $\gamma$ while keeping $\varphi$ and $\omega$ the same as that in **(a)**. **(d)** The SHG patterns for another ZnS NW that is oriented at $\varphi=58°$, $\gamma= -67°$ and $\omega= -10°$. **(e)** and **(f)** Theoretical calculations by just changing the value of $\gamma$ while keeping $\varphi$ and $\omega$ the same as that in **(d)**.

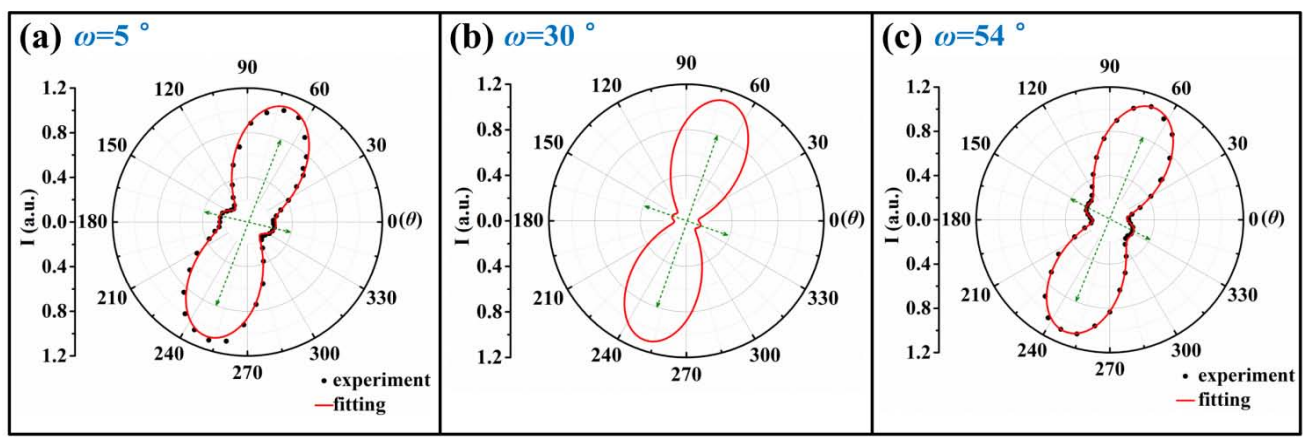

**Figure 4.** Polarization-dependent SHG patterns in different single ZnS NWs. The crystal orientations of ZnS NWs are determined to be (**a**) $\varphi=76°$, $\gamma=70°$, $\omega=5°$ and (**c**) $\varphi=76°$, $\gamma=68°$, $\omega=54°$. (**b**) Theoretical calculations by just changing the value of $\omega$ to 30° while keeping $\varphi$ and $\gamma$ the same as that in (**a**).

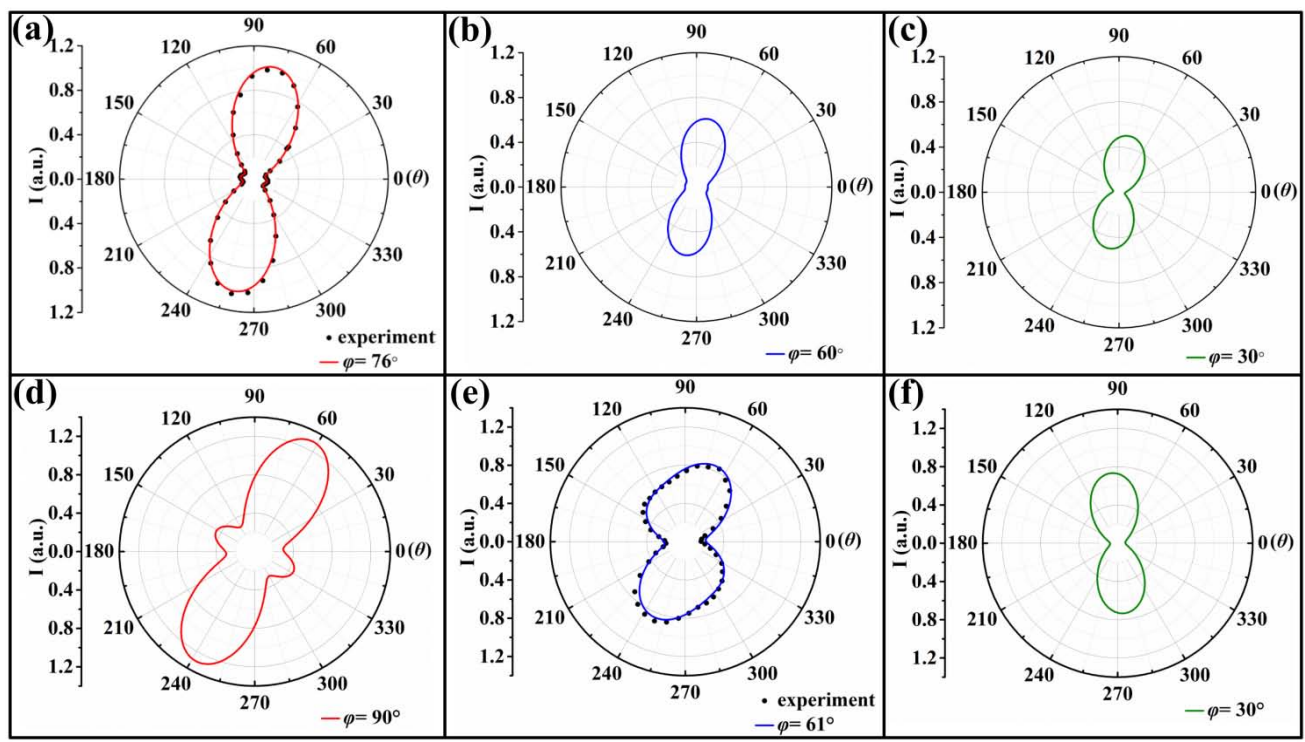

**Figure 5.** Polar plots of the measured SHG intensity as a function of the polarization angle $\theta$. The black dots present the experimental data, and the solid curves indicate the theoretical fits. (**a**) The SHG patterns for a single ZnS NW that has a crystal orientation of $\varphi=76°$, $\gamma=80°$, $\omega=30°$. The small

protuberances arise exactly along the direction perpendicular to the long-axis of two-lobes. (**b**) and (**c**) the related theoretical calculations by just changing the value of $\varphi$ while keeping $\gamma$ and $\omega$ the same as that in (**a**). (**e**) The SHG patterns for another ZnS NW that is oriented at $\varphi=63º$, $\gamma=62º$, $\omega=71º$. (**d**) and (**f**) Theoretical calculations by just changing the value of $\varphi$ while keeping $\gamma$ and $\omega$ the same as that in (**e**).